\documentclass[useAMS,usenatbib]{mnras}
\usepackage{graphicx}
\usepackage{amssymb}
   \title{New white dwarf and subdwarf stars in the Sloan Digital Sky Survey Data Release 12}

   \author[Kepler et al.]{S. O. Kepler$^{1}$\thanks{kepler@if.ufrgs.br},
		I. Pelisoli$^{1}$,
		D. Koester$^{2}$,
		G. Ourique$^{1}$,
		A. D. Romero$^{1}$,
\newauthor
		N. Reindl$^3$,
	        S. J. Kleinman$^4$,
		D. J. Eisenstein$^5$,
                A. D. M. Valois$^{1}$,
                L. A. Amaral$^{1}$\\
$^{1}$Instituto de F\'{\i}sica, Universidade Federal do Rio Grande do Sul,
              91501-900  Porto-Alegre, RS, Brazil\\
$^{2}$Institut f\"ur Theoretische Physik und Astrophysik, Universit\"at Kiel, 24098 Kiel, Germany\\
$^3$Institute for Astronomy and Astrophysics, Kepler Center for Astro and Particle Physics, Eberhard Karls University,\\
Sand 1, 72076, T\"ubingen, Germany\\
$^4$Gemini Observatory, Hilo, Hawaii, 96720, USA\\
$^5$Harvard Smithsonian Center for Astrophysics, 60 Garden St., MS \#20, Cambridge, MA 02138, USA\\
}
\begin{document}
\date{Accepted 2015 October 27.  Received 2015 October 26; in original form 2015 August 2}

\pagerange{\pageref{firstpage}--\pageref{lastpage}} \pubyear{2015}
   \maketitle

\label{firstpage}
  \begin{abstract}

We report the discovery of $6\,576$ new spectroscopically confirmed
white dwarf and subdwarf stars in the Sloan Digital Sky Survey Data 
Release 12. We obtain $T_\mathrm{eff}$, log g and mass for hydrogen 
atmosphere white dwarf stars (DAs) and 
helium atmosphere white dwarf
stars (DBs), estimate the calcium/helium abundances for the white
dwarf stars with metallic lines (DZs) and carbon/helium for carbon
dominated spectra DQs. 
We found  
one central star of a planetary nebula,
one ultra-compact helium binary (AM CVn),
one oxygen line dominated white dwarf,
15 hot DO/PG1159s, 
12 new cataclysmic variables,
36 magnetic white dwarf stars, 
54 DQs, 
115 helium dominated white dwarfs, 
148 white dwarf+main sequence star binaries,
236 metal polluted white dwarfs,
300 continuum spectra DCs, 
230 hot subdwarfs, 
$2\,936$ new hydrogen dominated white dwarf stars,
and $2\,675$ cool hydrogen dominated subdwarf stars.
We calculate the mass distribution of all 5883 DAs with S/N$\geq 15$ in DR12, including the ones in DR7 and DR10,
with an average S/N=26,
corrected to the 3D convection scale,
and also the distribution after correcting
for the observed volume, using $1/V_\mathrm{max}$. 
\end{abstract}


\section{Introduction}
White dwarf stars are the end product of evolution of all stars with progenitor masses below
7--10.6~$M_{\odot}$, depending on metallicity \citep[e.g][]{Ibeling13,Doherty14,Woosley15}, which corresponds to over 97\% of the 
total number of stars. Most white dwarfs do not generate energy from nuclear fusion, but radiate due to residual gravitational
contraction. Because of the degenerate equation of state, this is accompanied by a loss of thermal energy instead of increase
as in the case of ideal gases; the evolution of white dwarfs is therefore often simply described as cooling. The radius of a white
dwarf star is of the same order of the Earth's radius, which implies that they have small surface area, resulting in very large
cooling times (it takes approximately $10^{10}$ years for the effective temperature of a normal mass white dwarf to decrease
from $100\,000$~K to near $5\,000$~K). Consequently, 
the cool ones are among the oldest objects in the Galaxy. Therefore, studying white dwarfs is extremely important to 
comprehend the processes of stellar formation and evolution in the Milky Way \citep[e.g.][]{winget1987, BSL, Liebert2005, 
moehler2008, tremblay2014}.

The number of known white dwarf stars is increasing fast thanks to the Sloan Digital Sky Survey (SDSS). The first full white dwarf catalogue from SDSS data \citep{kle04} was based on SDSS Data Release 1 \citep[DR1,][]{dr1}. Using data from the SDSS Data Release 4 \citep[DR4,][]{dr4}, \citet{eis06} roughly doubled the number of spectroscopically confirmed white dwarf stars. In the white dwarf catalogue based on the SDSS Data Release~7 \citep[DR7,][]{rdr7}, \citet{dr7} increased the total number of white dwarf stars by more than a factor of two compared to the catalogue based on DR4 data. They also (re)analysed all stars from previous releases. Over $8\,000$ new spectroscopically confirmed white dwarf stars were reported by \citet{Kepler15} in the analysis of SDSS Data Release 10 \citep[DR10,][]{dr10}. They also improved the candidate selection compared to previous catalogues, implementing an automated search algorithm to search objects which were missed by the other selection criteria. It was also the first white dwarf catalogue based on SDSS data to fit not only DA and DB stars, but also DZ, DQ, and DA+MS pairs. We continue such detailed analysis here with SDSS Data Release 12 \citep[DR12,][]{dr12}. More details concerning the previous catalogues are presented in Table \ref{catalogues}.

Although the SDSS increased the number of spectroscopically-confirmed
white dwarf stars more than an order of magnitude prior to the SDSS,
the SDSS sample is far from complete.
Target selection considerations of the original SDSS (up to DR8)
implied that white dwarf selection for spectroscopy was incomplete
\citep[e.g.][]{Fusillo15}.
In the SDSS DR12, the ancillary target programme 42 \citep{Dawson13} obtained the spectra of an additional $762$ colour selected white dwarf candidates
that were missed by prior SDSS spectroscopic surveys, i.e., up to DR10. Here, we report on our search for new white dwarfs from the SDSS Data Release 12 \citep{dr12}, which in total observed photometrically one third of the celestial sphere and obtained  4.3 million useful optical spectra. Our catalogue does not include stars reported in the earlier catalogues,
except for classification corrections.

\begin{table}
\begin{center}
\caption{\label{catalogues}Number of objects and the main classifications in the previous white dwarf catalogues published based on SDSS data releases.}
\begin{tabular}{ccc}
{Catalogue} &  {Objects} & {Main classifications}\\
\\
DR1$^a$  & $2\,551$ WDs  & $1\,888$ DA\\
         & $240$ sds     & $171$ DB\\
\\
DR4$^b$  & $9\,316$ WDs  & $8\,000$ DA\\
         & $928$ sds     & $731$ DB\\
\\
DR7$^c$  & $19\,713$ WDs & $12\,831$ DA\\
         &               & $922$ DB\\
\\
DR10$^d$ & $8\,441$ WDs  & $6\,887$ DA\\
         & $647$ sds     & $450$ DB\\
\end{tabular}
$^a${\citet{kle04}}.
$^b${\citet{eis06}}.
$^c${\citet{dr7}, includes the (re)analysis of stars from previous releases, but does not include subdwarfs.}.
$^d${\citet{Kepler15}}.
\end{center}
\end{table}
\section{Target selection}

Even though targeting in SDSS produced the largest spectroscopic sample of white dwarfs, much of SDSS I and II white dwarf targeting required that the objects be unblended, which caused many bright white dwarfs to be skipped (for a detailed discussion, see Section 5.6 of Eisenstein et al. 2006). The BOSS ancillary targeting programme \citep{Dawson13} relaxed this requirement and imposed colour cuts to focus on warm and hot white dwarfs. Importantly, the BOSS spectral range extends further into the UV, covering from 3\,610~\AA\ to 10\,140~\AA, with spectral resolution 1560-2270 in the blue channel, and 1850-2650 in the red channel \citep{Smee13}, allowing full coverage of the Balmer lines.

The targeted white dwarfs in SDSS-III were required to be point sources with clean photometry, and to have USNO-B Catalog counterparts \citep{Monet03}. They were also restricted to regions inside the DR7 imaging footprint and required to have colours within the ranges
$g < 19.2$,
$(u-r) < 0.4$,
$-1 < (u-g) < 0.3$,
$-1 < (g-r) < 0.5$, 
and to have low Galactic extinction $A_r < 0.5$~mag. Additionally, targets that did not have $(u-r) < -0.1$ and $(g-r) < -0.1$ were required to have USNO proper motions larger than 2 arcsec per century. Objects satisfying the selection criteria that had not been observed previously by the SDSS ({\small ANC~42}) were denoted by the {\small WHITEDWARF\_NEW} target flag, while those with prior SDSS white dwarf photometric classification ({\small ANC~43}) are assigned the {\small WHITEDWARF\_SDSS} flag. Some of the latter were re-observed with BOSS in order to obtain the extended wavelength coverage that the BOSS spectrograph offers. The targeting colour selection included DA stars with temperatures above $\sim 14\,000$~K, helium atmosphere white dwarfs above $\sim 8\,000$~K, as well as many rarer classes of white dwarfs. Hot subdwarfs (sdB and sdO) were targeted as well.

Our selection of white dwarf candidates among DR12 objects was similar to that reported for DR10 \citep{Kepler15}. We did not restrict our sample by magnitude, but by S/N$\geq 3$. In addition to the 762 targeted white dwarf candidates after DR10 by {\small ANC~42}, we selected the spectra of any object classified by the {\small ELODIE} pipeline \citep{Bolton12} as a white dwarf, which returned 35\,708 spectra, an O, B or A star, which returned another 144\,471 spectra. Our general colour selection from \citet{dr7}, which takes into account that SDSS multi-colour imaging separates hot white dwarf and subdwarf stars from the bulk of the stellar and quasar loci in colour-colour space \citep{Harris03}, returned 68\,836 new spectra, from which we identified another 2\,092 white dwarfs, 79 subdwarfs, 36 cataclysmic variables, and 3 PG~1159. Most of these spectra were overlapping with the {\small ELODIE} selections.

We also used an automated search algorithm which assumes that the spectra of two objects with the same composition, effective temperature and surface gravity differ only in flux, due to different distances, and slope, because of reddening and calibration issues. This algorithm determines a polynomial of order between zero and two which minimizes the difference between each spectrum and a sample of models, allowing the determination of the most likely spectral class of each object. Running this search over the whole 4.5 million spectra from DR~12 recovered more than $80\%$ of our sample and found $400$ white dwarf stars missed by previous searches.

\section{Data Analysis}

The data analysed here were reduced by the {\small v$5\_0\_7$} spectroscopic reduction pipeline of \citet{Bolton12}. After visual identification of the spectra as a probable white dwarf, we fitted the optical spectra to DA and DB local thermodynamic equilibrium (LTE) grids of synthetic non-magnetic spectra derived from model atmospheres \citep{Koester10}. The DA model grid uses the ML2$/\alpha=0.6$ approximation, and for the DBs we use the ML2/$\alpha=1.25$ approximation, to be consistent with \citet{dr7} and \citet{Kepler15}. Our DA grid extends up to $T_\mathrm{eff}=100\,000$~K, 
but NLTE effects are not included. \citet{Napiwotzki97} concluded pure hydrogen atmospheres of DA white dwarfs are well represented by LTE calculations for effective temperatures up to 80\,000~K, but when traces of helium are present, non-local thermodynamic equilibrium (NLTE) effects on the Balmer lines occur, down to effective temperatures of 40\,000~K. 
\citet{Napiwotzki97} concluded LTE models should exclude
traces of helium for the analysis of DA white dwarfs. 
We fitted the spectral lines and photometry separately \citep{Koester10}, selecting between the hot and cool solutions using photometry as an indicator.

Of the $762$ objects targeted specifically as new white dwarf spectra by BOSS as Ancillary Programme 42, 19 were not identified as white dwarfs or subdwarfs by us. 
Of the Ancillary Program 43 of {\small WHITE\-DWARF\_SDSS} already observed, 5 in 80 colour selected stars are in fact quasars. 
\citet{Fusillo15} reports that only 40\% of their SDSS colour selected sample with high probability of being a white dwarf has spectra obtained by SDSS.

The SDSS spectra we classified as white dwarfs or subdwarfs have a g-band signal--to--noise ratio 85$\geq$S/N(g)$\geq$3, with an average of 12. The lowest S/N in the g-band occurs for stars cooler than $7\,000$~K, but they have significant S/N in the red part of the spectrum.

We include in our tables the new proper motion determinations of \citet{Munn14} and use them to separate DCs from BL Lac spectra.
We applied complete, consistent human identifications of each candidate white dwarf spectrum.

\subsection{Spectral Classification}

Because we are interested in obtaining accurate mass distributions for
our DA and DB stars, we were conservative in labelling a spectrum as a clean
DA or DB,  adding additional subtypes
and uncertainty notations (:) if we saw signs of other elements, companions, or
magnetic fields in the spectra.  While some of our mixed white dwarf
subtypes would probably be identified as clean DAs or DBs with better
signal-to-noise spectra,  few of our identified clean DAs or DBs would
likely be found to have additional spectral features within our detection
limit.

We looked for the following features to aid in the
classification for each specified white dwarf subtype:

\begin{itemize}
\item Balmer lines --- normally broad and with a steep Balmer decrement
[DA but also DAB, DBA, DZA, and subdwarfs]
\item HeI $4\,471$\AA\ [DB, subdwarfs]
\item HeII $4\,686$\AA\ [DO, PG1159, sdO]
\item C2 Swan band or atomic CI lines [DQ]
\item CaII H \& K  [DZ, DAZ, DBZ]
\item CII $4\,367$\AA\ [HotDQ]
\item Zeeman splitting [magnetic white dwarfs]
\item featureless spectrum with significant proper motion [DC]
\item flux increasing in the red [binary, most probably M companion]
\item OI $6\,158$\AA\ [Dox]
\end{itemize}

We also found 8 of stars to have an extremely
steep Balmer decrement (i.e. only a broad H$\alpha$ and sometimes
H$\beta$ is observed while the other lines are absent) that could not
be fit with a pure hydrogen grid, or indicated extremely high gravities. 
We find that these objects are
best explained as helium-rich DAs, and denote them DA-He.

We finally note that the white dwarf colour space also contains
many hot subdwarfs. It is difficult
to distinguish a low mass white dwarf from a subdwarf,
as they are both dominated by hydrogen lines and the
small differences in surface gravity cannot be spotted by
visual inspection alone. We therefore extended the model grid
to $\log g=5.00$ for $T_\mathrm{eff} \geq 25\,000$~K,
and $\log g=3.75$ for $T_\mathrm{eff} < 25\,000$~K,
to separate
white dwarfs ($\log g \geq 6.5$), subdwarfs ($6.5 > \log g \geq 5.5$) and
main sequence stars ($\log g \leq 4.75$)
(see section \ref{section:masses} and \ref{section:sub}),
but we caution that the differences in the line widths for DAs cooler than $\simeq 8000$~K
and hotter than $\simeq 30\,000$~K are minor, with changing gravity. 
We use sdA to denote spectra with $6.5 > \log g \geq 5.5$ and $T_\mathrm{eff} \leq 20\,000$~K.
Table~\ref{tb:ids} lists
the number of each type of white dwarf and subdwarf stars we identified.  

\begin{table}
\caption{\label{tb:ids}Numbers of newly identified stars by type.}
\begin{tabular}{rl}
{No. of Stars} &  {Type}\\
$2\,675$ &  sdA\\
$1\,964$ &  DA$^a$\\
300 & DC\\
236 & DZ\\
183 & sdB\\
104 & WD+MS$^b$\\
66 & DB\\
71 & DAZ\\
54 & DQ\\
47 & sdO\\
27 & DBA \\
28 & DAH\\
14 & DO/PG~1159\\
12 & CV\\
7 & DZH\\
6  & DAO\\
3 & DAB\\
2 & DBH\\
1 & DBZ\\
1 & Dox\\
1 & AM CVn (SDSS~J131954.47+591514.84)\\ 
1 & CSPN (SDSS~J141621.94+135224.2\\
\end{tabular}
$^a${Pure and certain DAs.}
$^{b}${These spectra show both a white dwarf star and a
companion, non-white dwarf spectrum, usually a main sequence M star.}

\end{table}

As an independent check, and to be consistent with the earlier 
SDSS white catalogues, we also fitted all DA and DB white
dwarf spectra and colours with the {\small AUTOFIT}
code described in \citet{kle04}, \citet{eis06} and \citet{dr7}.
{\small AUTOFIT} fits only clean DA and DB models.
In addition to the best
fitting model parameters, it also outputs 
flags for other features noted in the spectrum, like a possible dM companion.
These fits include SDSS imaging photometry and allow
for refluxing of the models by a low-order polynomial to incorporate
effects of unknown reddening and spectrophotometric flux calibration errors.

\section{Results}

\subsection{Masses\label{section:masses}}

\citet{dr7} limited the white dwarf classification to surface gravity $\log g \ge 6.5$.
At the cool end of our sample, $\log g=6.5$ corresponds to a mass around $0.2~M_\odot$, well below the single mass
evolution in the lifetime of the Universe. The He-core white dwarf stars in the mass range $0.2-0.45~M_\odot$, 
referred to as low-mass white dwarfs, are usually found in close binaries, often double degenerate systems \citep{Marsh1995}, being most likely a product of interacting binary stars evolution. 
More than 70\% of those studied by \citet{Brown11} with masses below $0.45~M_\odot$ and all but a few with masses below $0.3~M_\odot$ show velocity variations \citep{Brown13, Gianninas14}.
\citet{Kilic2007} suggests single low-mass white dwarfs result from the evolution of old metal-rich stars that truncate 
evolution before the helium flash due to severe mass loss. They also conclude all white
dwarfs with masses below $\simeq 0.3~M_\odot$ must be a product of binary
star evolution involving interaction between the components, otherwise
the lifetime of the progenitor on the main sequence would be
larger than the age of the Universe. 

DA white dwarf stars with masses $M\leq 0.45~M_\odot$
and $T_\mathrm{eff} < 20\,000$~K are Low Mass and Extremely Low Mass (ELM) as found by
\citet{Brown10}, \citet{Brown11}, \citet{Brown12a}, \citet{Brown12b}, \citet{Brown13}, and \citet{Gianninas14}.
\citet{Hermes12}, \citet{Hermes13a}, \citet{Hermes13b}, and \citet{Bell15} found pulsations 
in six of these ELMs,
similar to the pulsations seen in
DAVs (ZZ~Ceti stars), as described in \citet{VanGrootel13}.
\citet{Maxted14a} found 17 pre-ELMs,
i.e., helium white dwarf precursors,
and
\citet{Maxted14b} report pulsations in one of them.
Pulsations are an important tool to study the stellar interior, and
\citet{Corsico14a}, \citet{Corsico14b}, \citet{Corsico15}, \citet{Istrate14a}, and \citet{Istrate15} report on theoretical models and pulsations of ELMs. 

We classified as DAs those with $\log g\geq 6.5$ as in \citet{dr7},
and sdAs those with $6.5>\log g\geq 5.5$ when $T_\mathrm{eff}\leq 20\,000$~K (see section~\ref{section:sub}).
Low metallicity main sequence stars have an upper limit to $\log g\lesssim 4.75$. To select the low $\log g$ limit,
we use an external, systematic uncertainty 
in our surface gravity determinations of $3\sigma(\log g)=3\times 0.25$,
around 15$\times$ our average internal fitting uncertainty.

\begin{figure}
   \centering
   \includegraphics[width=\linewidth]{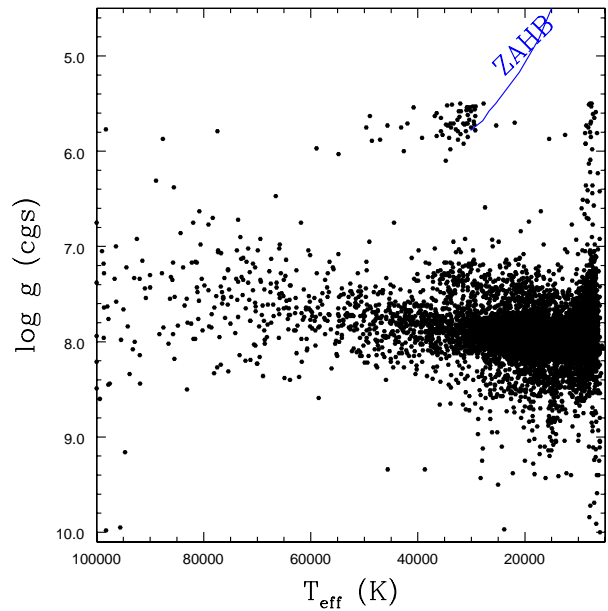}
      \caption{
Surface gravity ($\log g$) and effective temperature ($T_{\rm eff}$) estimated for all DA white dwarf stars in DR7 to DR12 with
spectral S/N$_g\geq 15$, after applying three-dimensional convection atmospheric model corrections
from \citet{Tremblay13}. 
The Zero Age Horizontal Branch (ZAHB) plotted was calculated with solar composition models.
It indicates the highest possible surface gravity for a hot subdwarf. Stars with 
$T_\mathrm{eff}\leq 45\,000$~K and
smaller surface gravities 
than the ZAHB are sdBs.}
         \label{dr712loggDA}
   \end{figure}

Fig.~\ref{dr712loggDA} 
shows surface gravity, $\log g$, as a
function of the effective temperature $T_\mathrm{eff}$, estimated for all $5\,884$
DAs spectroscopically identified in DR7 to DR12 with SDSS spectral S/N$\geq 15$. 
We include corrections to $T_\mathrm{eff}$ and $\log g$ based on three--dimensional convection
calculations from \citet{Tremblay13}.

We use the
mass--radius relations of \citet{Renedo10} and \citet{Romero15}
for
carbon--oxygen  DA  white dwarfs  with  solar metallicities
to calculate the mass of our identified DA stars from the
$T_\mathrm{eff}$ and  $\log g$ values obtained from our fits,
after correcting to 3D convection.
These mass--radius
relations are based on full evolutionary calculations appropriate for
the study of hydrogen-rich DA  white dwarfs that take into account the
whole evolution  of progenitor stars. The sequences are computed from the  zero--age main sequence,
through  the hydrogen and helium central burning stages, thermal pulsations
and mass--loss in the asymptotic giant branch phase and finally the planetary
nebula domain. The white dwarf masses for the resulting sequences
range from 0.525 to 1.024 $M_{\odot}$, covering  the stellar
mass range for C--O core DAs. For high--gravity white dwarf
stars, we used the mass--radius relations for O--Ne core
white dwarfs given in \citet{Althaus05}
in the mass range
from 1.06 to 1.30 $M_{\odot}$ with a step of 0.02 $M_{\odot}$. For
the low--gravity white dwarf and cool subdwarf stars, we used the evolutionary
calculations of \citet{Althaus13} for helium--core white dwarfs with
stellar mass between 0.155 to 0.435 $M_{\odot}$, supplemented by
sequences of 0.452 and 0.521 $M_{\odot}$ calculated in \citet{Althaus09a}.

The spectra we classified as DBs belong to 116 stars.
27 of these are DBAs, one is a DBZ (SDSS~J122649.96+444513.59), and 8 are DB+M. To calculate the DB white dwarf masses in the catalogue, we relied on the evolutionary calculations
of hydrogen--deficient white dwarf stars with stellar masses between
0.515 and 0.870 $M_{\odot}$ computed by \citet{Althaus09b}. These
sequences have  been derived from the born--again episode responsible
for the  hydrogen deficient white dwarfs. For  high-- and low--gravity DBs, we used
the same O-Ne and helium evolutionary sequences described before.

To calculate a reliable mass distribution for DAs, we selected only the S/N$\geq 15$
spectra with temperatures well fit by our models.
Including the DAs from DR7 \citep{dr7} and DR10 \citep{Kepler15}, we classified a total of $5\,884$
spectra as clean DAs
with S/N $\geq 15$, with a mean S/N=$26\pm 11$.
Table~\ref{mass} presents the mean masses for different signal-to-noise limits.
\citet{Gianninas05} estimate the increase of the uncertainty in the surface gravity from $\Delta \log g\simeq 0.06$~dex to 
$\Delta \log g\simeq 0.25$~dex when the S/N decreases from 50 to 15.
\citet{Genest14} conclude there appears to be a small residual zero point offset in the absolute fluxes of SDSS spectra.
If the differences in the mean masses with S/N are not due to systematic (not random) effects, it could be the reflection
of different populations, as faint stars perpendicular to the disk of the Galaxy could
have different metallicities, and therefore different star formation mass functions and different Initial-to-Final-Mass
relations \citep{Romero15}.

\begin{table}
\begin{center}
\begin{tabular}{crc}
S/N$_g$&N&$\langle M_{\mathrm{DA}} \rangle $\cr
&&$(M_\odot)$\cr
15&5884&$0.608\pm 0.002$\cr
25& 2591&$0.620\pm 0.002$\cr
50& 265&$0.644\pm 0.008$\cr
\end{tabular}
\caption{Mean masses for all DAs, corrected to 3D convection.\label{mass}}
\end{center}
\end{table}
The mean masses estimated in our DR7 to DR12 sample are smaller than those obtained by
\citet{Kepler15},
even with the use of the 3D corrections for the whole sample.

\begin{figure}
   \centering
   \includegraphics[width=\linewidth]{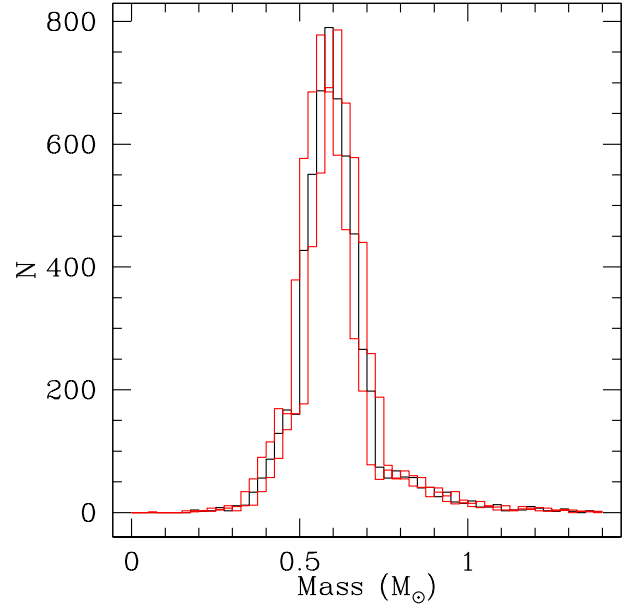}
      \caption{
Histogram for the mass distribution of 5884 S/N$\geq$15 DAs versus mass, for $\log g$ corrected to three-dimensional convection models using
the corrections reported in \citet{Tremblay13}.
The colored lines show the $-1\sigma$ and $+1\sigma$ uncertainties.
              }
         \label{masshist}
   \end{figure}

\begin{figure}
   \centering
   \includegraphics[width=\linewidth]{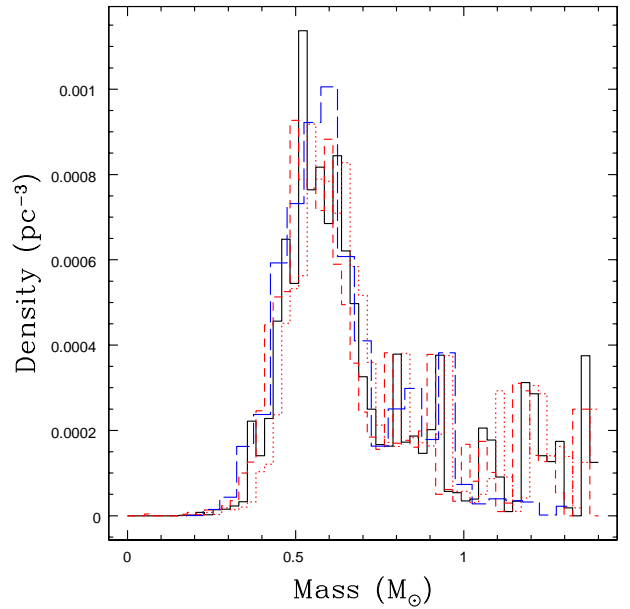}
      \caption{
Histogram for the density distribution of S/N$\geq$15 DAs versus mass, for $\log g$ corrected to three-dimensional convection models,
after correcting by the observed volume and by 40\% completeness from \citet{Fusillo15}.
The colored lines show the $-1\sigma$ and $+1\sigma$ uncertainties.
The long dashed (blue) histogram is
the one from \citet{Rebassa15}, limited to $M_{bol}>13$.
              }
         \label{masshistv}
   \end{figure}

Fig.~\ref{masshist} shows the mass histogram for the 5\,884 DAs with S/N $\geq$ 15
and Fig.~\ref{masshistv} shows the mass distribution after correcting by the observed volume,
following \citet{vmax,Schmidt75}, \citet{Green80}, \citet{Stobie89}, \citet{liebert03}, \citet{Kepler07}, \citet{limoges10} and \citet{Rebassa15}.
This correction takes into account the shape of the galactic disk, assuming an scale height of
250~pc, minimum ($g\simeq 14.5$) and maximum ($g=19$) magnitudes, for a complete sample. 
\citet{Green80} propose completeness can be estimated
from $\langle V/V_{max}\rangle$, which is equal to 0.48 in our sample, close to the expected value of 0.50.

\citet{Rebassa15} limit their sample to bolometric magnitude
$M_{bol}\leq 13$, because \citet{Fusillo15} estimates completeness of
40\% down to this magnitude. Such bolometric magnitude corresponds to $T_\mathrm{eff}\gtrsim 8500$~K around masses $0.6~M_\odot$,
and to $T_\mathrm{eff}\gtrsim 10\,000$~K around masses $1.0~M_\odot$.
We do not limit our sample to $M_{bol}\leq 13$.
We find 94 DA white dwarf stars with masses above $1.0~M_\odot$ and S/N$\geq 15$, and applying the
volume correction to them, find a lower limit to their density of $0.000\,026\,M_\odot$~pc$^{-3}$.
20 of these have $M_{bol}>13$.
We did not apply any completeness correction by proper motion 
\citep[e.g.][]{Lam15}
because we did not apply a consistent limit on the proper motion.
The distribution for masses above the main peak around $0.6~M_\odot$ is
significantly uneven, possibly the outcome of distinct formation
mechanisms, including single star formation, accretion and mergers.

The DB mass distribution obtained from models including hydrogen contamination, is discussed in \citet{Koester15}.
{\bf As our temperatures and surface gravities were estimated with pure DB models, while those of
\citet{Koester15}
include Hydrogen contamination, their values are more accurate.}

\subsection{Magnetic Fields and Zeeman Splittings}
When examining each white dwarf candidate spectrum by eye, we
found 36 stars with Zeeman splittings indicating magnetic fields
above 2~MG --- the limit where the line splitting becomes too small
to be identified at the SDSS spectral resolution. This number is similar to
our findings reported for DR7 in \citet{kepler13} and DR10 \citep{Kepler15}.

If the line splitting and magnetic fields were not recognized, the spectral fittings of DA and DB models would have rendered
too high $\log g$ determinations due to magnetic broadening being misinterpreted as pressure broadening.

We also identified seven cool DZH (Table~\ref{dzh}), similar to those identified by \citet{Hollands15}.
\begin{table}
\begin{center}
\caption{Magnetic field for DZHs\label{dzh}}
\begin{tabular}{lccc}
SDSS~J&Plate-MJD-Fiber&B    &$\sigma$(B)\cr
      &     &(MG) &(MG)\cr
003708.42-052532.80&7039-56572-0140&7.2&0.2\cr
010728.47+265019.94&6255-56240-0896&3.4&0.1\cr
110644.27+673708.64&7111-56741-0676&3.3&0.1\cr
111330.27+275131.41&6435-56341-0036&3.0&0.1\cr
114333.46+661532.01&7114-56748-0973&9.0&1.5\cr
225448.83+303107.15&6507-56478-0276&2.5&0.1\cr
233056.81+295652.68&6501-56563-0406&3.4&0.3\cr
\end{tabular}
\end{center}
\end{table}

\begin{figure}
   \centering
   \includegraphics[width=\linewidth]{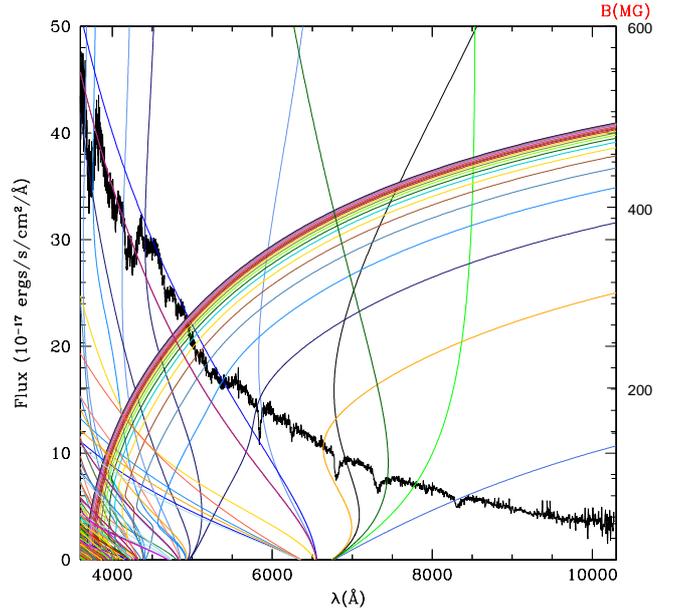}
      \caption{
Observed spectrum of the DAH, SDSS~J112148.77+103934.1 with $g=17.98 \pm 0.03$ and $B\simeq 309$~MG.
The coloured lines indicate the
positions of each theoretical Zeeman split Balmer line subcomponent, assuming a dipole magnetic field of strength indicated in the right axis.
Even low fields produce large splittings of the
higher Balmer lines.  The theoretical models are from 
\citet{Schimeczek13,Schimeczek14,Christopher14}.
              }
         \label{dr12dah}
   \end{figure}

We estimated the mean fields for the new DAHs following \citet{Kulebi} as being from 3~MG to 80~MG.
We caution that stars with large fields are difficult to identify because 
fields above around 30~MG, depending on effective temperature and
signal-to-noise, 
intermixes subcomponents between different hydrogen series components so much that it becomes difficult to identify the star as containing hydrogen at all,
and affect the colours significantly.
Additionally white dwarf stars with fields above 100~MG (see Fig.~\ref{dr12dah}) 
represent the intermediate regime in which the spectra have very few features,
except for a few stationary transitions that have similar wavelengths for a
reasonable distribution of magnetic fields over the surface of the star.

In \cite{dr7} and \cite{kepler13}, we mis-classified SDSS~J110539.77+250628.6, Plate-MJD-Fiber (P-M-F)=2212-53789-0201 and 
SDSS~J154012.08+290828.7, P-M-F=4722-55735-0206 as magnetic,
but they are in fact CVs. SDSS~J110539.77+250628.6 was identified as an AM~Her star by \citet{Liu12}. Here, we update
the identification of SDSS~J154012.08+290828.7 to a cataclysmic variable (CV), with a period around 0.1~d based on data from the Catalina Sky Survey \citep[CSS,][]{Drake09}.
We found another 14 cataclysmic variables based on seeing hydrogen and/or helium lines in emission.
Most are variable in the CSS.

\subsection{DCs and BL~Lac}
\noindent

Featureless optical spectra are the signature of DC white dwarfs, but also from extragalactic BL~Lac objects. BL~Lac objects
are strong sources of radio emission, while non-interacting DCs are not. DCs, if bright enough to be detected in all images,
generally have measurable proper motions, as their inherent dimness means they are relatively close to us.
To separate DCs form BL~Lacs, we searched for 1.4~GHz radio emission in the literature and looked for measured proper motions in
\citet{Munn14}.
We found 41 of our DC candidates were really BL Lac objects based on detectable radio emission.
{\bf We discarded the objects with radio emission,
as well as those with no radio emission and no proper motion.}


\subsection{DZs}
Of the new white dwarfs in our sample, 3\% have spectra with metal lines, probably
due to accretion of rocky material around the stars \citep[e.g.][]{Graham90, Jura03, Koester14}.
Calcium and magnesium in general have the strongest lines for white dwarfs at these temperatures.

We identified two DBZs as having unusual oxygen lines. 
SDSS~J124231.07+522626.6, {\small P-M-F} 6674-56416-0868, with $T_\mathrm{eff}=13\,000$~K,
was misclassified as an sdB from spectrum {\small P-M-F} 0885-52379-0112 in \citet{eis06},
and identified by us here as an oxygen-rich DBZ, possibly formed by accretion of an water rich asteroid as suggested by
\citet{Raddi15} and {\bf \citet{Farihi13}}.
SDSS~J123432.65+560643.1, spectrum {\small P-M-F} 6832-56426-0620,
was identified as DBZA in \citet{dr7} from {\small P-M-F} 1020-52721-0540. 
but is a DBZ.
We estimate $T_\mathrm{eff} = 12\,400\pm 120$~K, $\log g = 8.135\pm 0.065$.

\begin{figure}
   \centering
   \includegraphics[width=\linewidth]{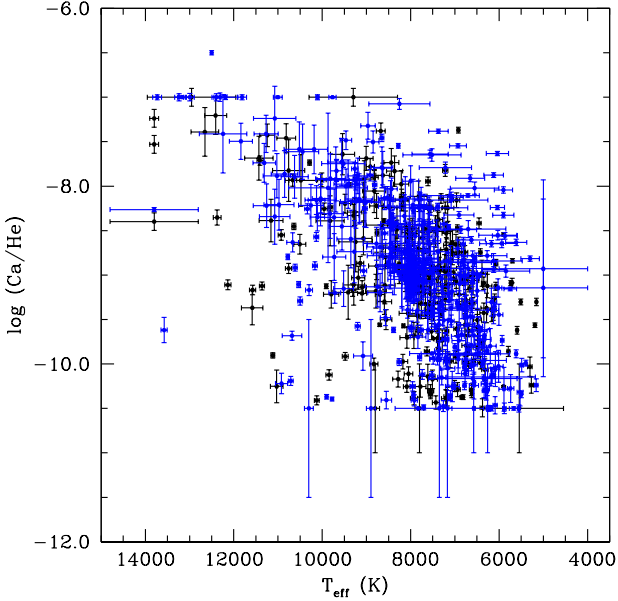}
      \caption{Calcium/Helium abundances estimated for DZs contained in
	      DR7 to DR12.
              }
         \label{dzs}
   \end{figure}

We fit the spectra of each of the 236 stars classified as DZs to
a grid of models with 
Mg, Ca and Fe ratios equal to the averages from the cool DZs in 
\citet{Koester11},
and Si added with the same abundance as Mg
\citep{Koester14}. 
These models have a fixed
surface gravity of $\log g=8.0$ as it is not possible to otherwise estimate it
from the spectra. 
The absolute values for log Ca/He range from -7.25 to -10.50.
Fig.~\ref{dzs} shows the calcium/helium abundance for the 246 DZs identified in DR12, in addition to those of DR7 and DR10.
There seems to be a decrease of Ca/He abundances at lower temperatures.
This trend might be explained if all stars had the same metal-rich material accretion rate, but the material becomes more diluted
at cooler temperatures due to the increasing convection layer size.

\subsection{DQs}
Only 0.7\% 
of the newly identified spectra
in our sample are dominated by carbon lines that are believed to be
due to dredge-up of carbon from the underlying carbon-oxygen core through the expanding He convection zone
\citep[e.g.][]{Koester82,Pelletier86,Koester06,Dufour07}.
These stars are in general cooler than $T_\mathrm{eff}=12\,000$~K.

\begin{figure}
   \centering
   \includegraphics[width=\linewidth]{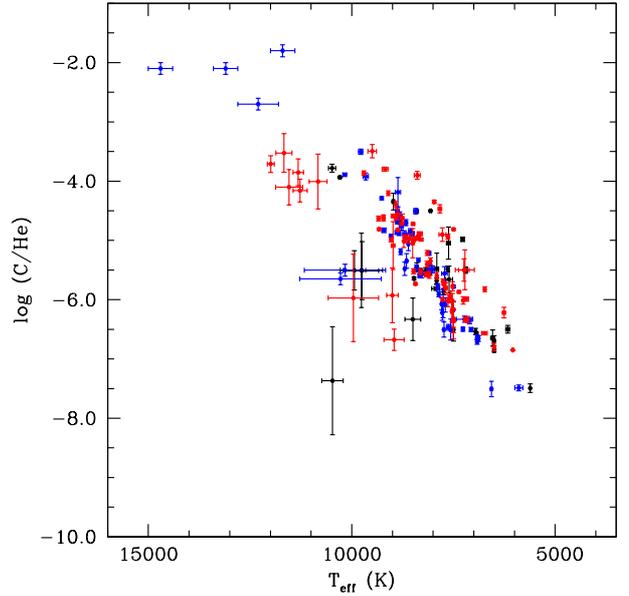}
      \caption{Carbon/Helium abundances estimated for DQs.
The decrease with decreasing temperature comes from the increase in transparency and deepening convection zone.
The darker points are the new DQs from DR12.
The lighter (red and blue) points are the results of our fits with the same models for the cool DQs in \citet{dr7} and \citet{Kepler15}.
              }
         \label{dqs}
   \end{figure}

We fitted the spectra of the stars (classified as cool DQs) to
a grid of models reported by 
\citet{Koester06}.
The models have a fixed
surface gravity of $\log g=8.0$ as it is not possible to otherwise estimate it
from the spectra. 
The absolute values for log C/He range from -8 to -4, and effective temperatures vary
from 13\,000~K to 4400~K.
Fig.~\ref{dqs} shows the carbon/helium abundance for the 54 new cool DQs identified here in addition to those from DR7 and DR10.
There is a decrease of C/He abundances at lower temperatures,
probably caused by the deepening of the convection zone, diluting any surface carbon.

\subsection{White dwarf-main sequence binaries}

We have identified 104 new white dwarfs that are part of apparent binary systems
containing main--sequence companions (WD-MS binaries).
The majority (96) of our new systems contain a DA white dwarf and a M dwarf
secondary star (DA+M).

\subsection{Subdwarfs\label{section:sub}}

Hot subdwarfs are core He burning stars.
Following \citet{Nemeth12}, \citet{Drilling13}, \citet{Nemeth14a}, and \citet{Nemeth14b},
we have classified stars with $\log g< 6.5$ and $45\,000~\mathrm{K} \geq T_\mathrm{eff} \geq 20\,000$~K as subdwarfs:
sdOs if He~II present and sdBs if not.
\citet{Nemeth14a} and \citet{Rauch14} discuss how the He abundances
typical for sdB stars affect the Non Local Thermodynamical Equilibrium (NLTE) atmosphere structure.
To a lower extent, CNO and Fe abundances are
also important in deriving accurate temperatures and gravities. 
Our determinations of $T_\mathrm{eff}$ and $\log g$ do not include
NLTE effects or mixed compositions, so they serve only as a rough estimate.
We classified 47 new sdOs and 183 new sdBs.

The Extreme Low Mass white dwarf catalog \citep{Gianninas15} lists 73 stars with $\log g \geq 4.8$, most with detected radial velocity variations
demonstrating they are in binary system.
We classified the hydrogen dominated spectra with $6.5 > \log g_\mathrm{sdA} \geqslant 5.5$ 
and $T_\mathrm{eff} \leq 20\,000$˜K as sdAs. These spectra look like main sequence low metallicity
A stars, but their estimated surface gravities with $\log g_\mathrm{sdA} \geqslant 5.5$ are at least $3\sigma$ (external) larger than those of main sequence stars $\log g_\mathrm{MS}< 4.75$
\citep{Heiter15}. We caution that the
spectral lines and colours used in our analysis are weakly dependent on surface gravity for $T_\mathrm{eff} \leq 9\,000$~K.
Even though a few of these stars have been classified previously as horizontal branch stars, to our knowledge, this is the first analysis with model spectra covering the range of surface gravities $3.75 \leqslant \log g \leqslant 10$.
Of these sdAs, 1275 have estimated proper motions larger than 5~mas/yr, and 476 larger than 10~mas/yr.
Because their spectra consists mainly of hydrogen lines, with cooler ones showing also Ca H\&K,
but no G- or CN-bands, we define their spectral types as sdA. We propose many of them are ELMs \citep{Corsico14a,Corsico14b,Corsico15,Istrate14a,Istrate14b,Istrate15}
but until
their binarity can be established \citep[e.g.][]{Gianninas15}, we classify only their spectral type.

\subsection{Noteworthy individual objects}

\begin{figure}
   \centering
   \includegraphics[width=\linewidth]{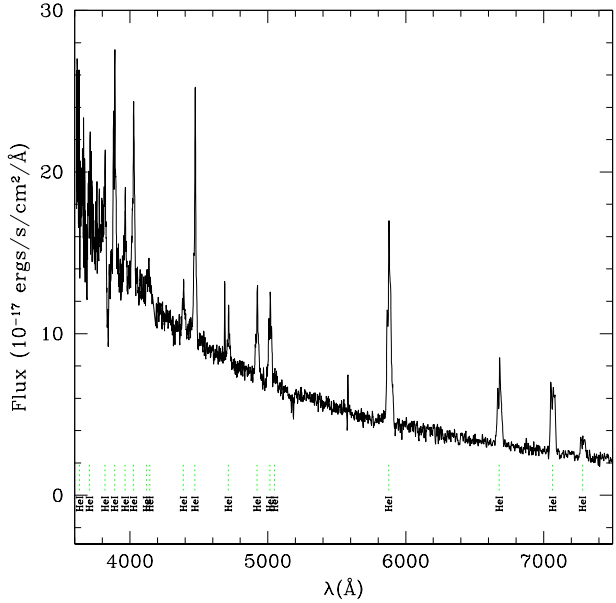}
      \caption{Spectra of the AM~CVn double degenerate SDSS~J131954.47+591514.84 ($g=19.106\pm 0.015$, proper motion = 33 mas/yr).
              }
         \label{amcvn}
   \end{figure}

Fig.~\ref{amcvn} shows the spectrum of the AM~CVn type ultra-compact
double degenerate binary, SDSS~J131954.47+591514.84, a new classification of a star with He-dominated atmosphere and He transfer. 
AM CVn objects are thought to be strong sources of gravitational waves \citep{nelemans2005}; 
however, only 
43 such objects are known \citep{Campbell15,Levitan15}.

\begin{figure}
   \centering
   \includegraphics[width=\linewidth]{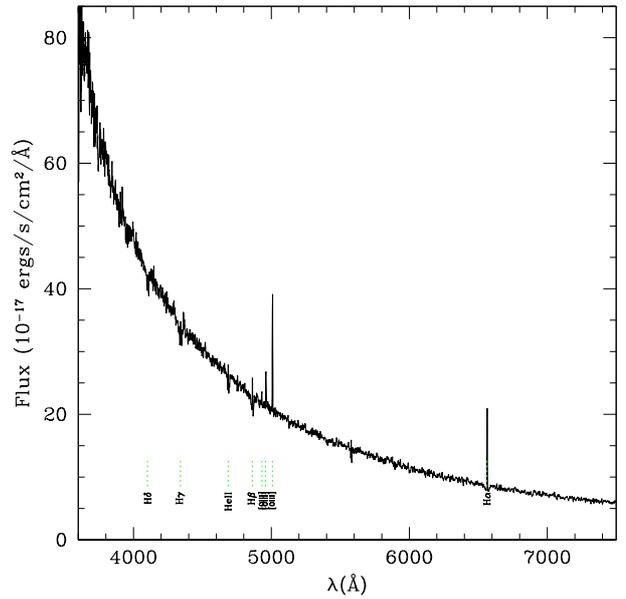}
\caption{Spectra of the CSPN SDSS~J141621.94+135224.20 ($g=18.202\pm 0.019$, proper motion = 12~mas/yr).
              }
         \label{cspn}
   \end{figure}

SDSS~J141621.94+135224.20 (spectra with P-M-F 5458-56011-0636, Fig.~\ref{cspn})
is a hot central star of a faint planetary nebula (CSPN) (PN~G003.3+66.1) and was misclassified by \citet{Fusillo15} as a cataclysmic variable. 
It is listed in the Southern H\,$\alpha$ Sky Survey Atlas, however we could not detect any planetary nebula in either the SDSS image or  WISE images. 
Thanks to its higher S/N, the new 
SDSS spectrum reveals the nebular emission lines of [OIII] $\lambda\lambda$ 4931, 4956, 5007~\AA, 
which we have now identified for the first time in this star.
The lack of He~I absorption lines indicates that the central star is hotter than 70\,kK. 
All photospheric absorption lines (H~$\delta$, H~$\gamma$, H~$\beta$, H~$\alpha$ as well as HeII $\lambda\lambda$~4686, 5412~\AA) show central emissions. 
They are likely nebular lines, however a photospheric contribution cannot be excluded for very hot central stars.

SDSS J103455.90+240905.75 (P-M-F 6439-56358-0445) was classified by \citet{Girven11} as DAB from spectra 2352-53770-0124,
SDSS J100015.28+240724.60 (P-M-F 6459-56273-0598) was classified as DB from spectra 2344-53740-0137, and
SDSS J101935.69+254103.04 (P-M-F 6465-56279-0808) was classified as DA from spectra 2349-53734-0523,
but the new higher S/N spectra shows they are DOs.


\vspace{0.5cm}
\noindent Table~\ref{tb:columns} lists the columns of data provided
in our electronic catalogue file, Table~\ref{dados}.
\begin{table*}
\caption{\label{tb:columns}Columns provided in data table, Table~\ref{dados}.}
\begin{tabular}{cll}
{Column No.} & {Heading} & {Description}
\\
1 & Name & SDSS object name (SDSS 2000J+) \\
2 & P-M-F & SDSS Plate number-Modified Julian date-Fiber\\
3 & SN\_g & SDSS g band signal to noise ratio \\
4 & u\_psf & SDSS u band PSF magnitude \\
5 & u\_err & SDSS u band uncertainty \\
6 & g\_psf & SDSS g band PSF magnitude \\
7 & g\_err & SDSS g band uncertainty \\
8 & r\_psf & SDSS r band PSF magnitude \\
9 & r\_err & SDSS r band uncertainty \\
10 & i\_psf & SDSS i band PSF magnitude \\
11 & i\_err & SDSS i band uncertainty \\
12 & z\_psf & SDSS z band PSF magnitude \\
13 & z\_err & SDSS z band uncertainty \\
14 & E(B-V) & color excess \\
15 & PM & USNO proper motion (mas yr$^{-1}$) \\
16 & l & galactic longitude (degrees) \\
17 & b & galactic latitude (degrees) \\
18 & T\_eff & $T_\mathrm{eff}$ (K) \\
19 & T\_err & $T_\mathrm{eff}$ uncertainty (K) \\
20 & log\_g & $\log{g}$ (cgs) \\
21 & log\_gerr & $\log{g}$ uncertainty (cgs)\\
22 & humanID & human classification\\
23 & T\_eff (3D) & $T_\mathrm{eff}$ for pure DAs and DBs or -log(Ca/He) for DZs or -log(C/He) for DQs\footnote{The temperatures and surface gravities are corrected to the three--dimensional
convection models of \citet{Tremblay13}. The Ca/He and C/He abundances, calculated assuming $\log g=8.0$, are indicated by -log(Ca/He) or - log(C/He).}\\
24 & T\_err  (3D)& $T_\mathrm{eff}$ uncertainty \\
25 & log\_g  (3D)& $\log{g}$ \\
26 & log\_gerr  (3D)& $\log{g}$ uncertainty\\
27 & Mass & calculated mass ($M_\odot$), corrected to 3D-convection\\
28 & Mass\_err & mass uncertainty ($M_\odot$), corrected to 3D-convection\\
\end{tabular}
\end{table*}

Table~\ref{old} lists 409 new classifications of stars already in Simbad (Strasbourg Astronomical Data Center), but for which new higher S/N spectra lead us to
a different classification.

\section{Conclusions and Discussion}

We have identified $6\,576$ new white dwarf and subdwarf stars in the DR~12 of the SDSS,
and estimated the masses for DAs and DBs, as well as the calcium contamination in DZs and carbon/helium abundances in DQs.
We were able to extend our identifications down to $T_\mathrm{eff}=5\,000$~K, although these are certainly not
complete, as we relied also on proper motion measurements to distinguish between cool DCs and BL~Lac objects.  Proper motions are
typically incomplete below $g\simeq 21$.
The resultant substantial increase in the number of spectroscopically confirmed white dwarfs is important because it
allows the discovery of more rare objects like massive white dwarfs, magnetic white dwarfs, and He-dominated objects with oxygen lines.
Extending the work of \citet{Kepler07} and \citet{Rebassa15}, we find 94 white dwarf stars with masses
above 1~$M_\odot$ and S/N$\geq$15. 
Their volume corrected distribution is inhomogeneous which, if confirmed, indicates
multiple formation processes, including mergers.
The volume-limited sample of white dwarfs within 40 pc by 
\citet{Limoges15} finds ~8\% (22/288) of the local sample of white dwarfs
have masses $M>1\,M_\odot$.

Massive white dwarfs are relevant both to the lower limit of core collapse SN and to white dwarf explosion
or merger as SN~Ia.
\citet{Nomoto13} estimates that the observed
$^{64}\mathrm{Zn}$ abundances provides an upper limit to the occurrence of exploding O-Ne-Mg cores 
at approximately 20\% of all core-collapse supernovae.
The existence of different types of SN~Ia indicates different types of progenitors do exist.

With our spectral model grid now extending from $3.75 \leq \log g \leq 10$, we identified 
2675 stars with hydrogen dominated spectra, and surface gravities $3-7\sigma$ larger than those of main sequence stars.
Time-series spectroscopy is necessary to check if they
are binaries, in order to establish what fraction of the sdA objects are ELM white dwarfs. If they were to have
main sequence radii, their distances would be tens of kiloparsecs outside the disk due to their distance moduli of
$15 \leq m-M \leq 20$.  The substantial fraction of these stars that have 
measured proper motions, if at large distances, would also be runaway stars or hypervelocity stars ($v>600$~km/s) \citep{Brown15}. 
The significant number
of these stars probably indicates Population II formation lead to a considerable ratio of binary stars.
\citet{DeRosa14} determine that $69\pm 7$\% of all A stars in the solar neighborhood are in binaries.
The pre-white dwarf ages of low metallicity stars with main sequence masses $0.9~M_\odot$ can amount to more than 8~Gyr,
so white dwarfs originating from binary interactions of low mass, low metallicity stars 
should still be visible as extremely low mass white dwarfs.
If the stars we classified as sdAs are in fact A-type main sequence stars, there is a large number of those at large 
distances from the galactic disk, and
the galactic formation model would have to account for the continuous formation of low metallicity stars,
perhaps from the continuous accretion of dwarf galaxies \citep{Camargo15,des15}.
Our analysis of their spatial distribution shows no concentrations.

\section*{Acknowledgments}

S.O.K.,
I.P.,
G.O.,
A.D.R., and A.D.M.V.
are sup\-por\-ted by CNPq-Brazil.
D.K. received support from programme Science without Borders, MCIT/MEC-Brazil.
N.R. is supported by the German Aerospace Center (DLR,  grant 05 OR 1507).
This research has made use of NASA's Astrophysics Data System.

Funding for SDSS-III has been provided by the Alfred P. Sloan Foundation, the Participating Institutions, 
the National Science Foundation, and the U.S. Department of Energy Office of Science. The SDSS-III web site is http://www.sdss3.org/.

SDSS-III is managed by the Astrophysical Research Consortium for the Participating Institutions of the SDSS-III Collaboration including the University of Arizona, the Brazilian Participation Group, Brookhaven National Laboratory, Carnegie Mellon University, University of Florida, the French Participation Group, the German Participation Group, Harvard University, the Instituto de Astrofisica de Canarias, the Michigan State/Notre Dame/JINA Participation Group, Johns Hopkins University, Lawrence Berkeley National Laboratory, Max Planck Institute for Astrophysics, Max Planck Institute for Extraterrestrial Physics, New Mexico State University, New York University, Ohio State University, Pennsylvania State University, University of Portsmouth, Princeton University, the Spanish Participation Group, University of Tokyo, University of Utah, Vanderbilt University, University of Virginia, University of Washington, and Yale University.

\begin{table*}
\tiny
\noindent
\begin{minipage}{\linewidth}
\caption{New White Dwarf Stars. Notes: 
P-M-F are the Plate-Modified Julian Date-Fiber number that designates an SDSS spectrum.
A {\bf:} designates an
uncertain classification. 
The columns are fully explained in Table~\ref{tb:columns}.
When $\sigma(\log g)=0.000$, we have assumed $\log g=8.0$, not fitted the surface gravity.
The full table is available on online, and at http://astro.if.ufrgs.br/keplerDR12.html.
\label{dados}}
\tiny
\begin{verbatim}
#SDSSJ              P-M-F           S/N u      su    g      sg    r      sr    i      si    z      sz    E(B-V) ppm  long  lat   sp   Teff  sT     logg slogg  Teff dTeff logg dlogg mass  dmass
#                                       (mag) (mag) (mag)   (mag) (mag) (mag) (mag)   (mag) (mag) (mag) (mag)  0.001"  (o)   (o)     (K)   (K)  (cgs)  (cgs) (K)  (K) (cgs)  (cgs) (Msun)  (Msun)
000000.46+174808.91 6207-56239-0156 005 21.460 0.125 20.966 0.033 20.992 0.045 21.063 0.068 21.211 0.322 0.028 024.4 106.0 -43.4 DA   09683 00132 7.945 0.174 09635 0132 7.680 0.170 0.459 0.067
000007.84+304606.35 7134-56566-0587 011 20.208 0.039 19.665 0.017 19.534 0.021 19.495 0.023 19.562 0.059 0.060 44.69 110.1 -30.8 DA:  07545 00064 7.616 0.117 07566 0064 7.470 0.120 0.366 0.042
000013.17-102750.57 7167-56604-0281 006 21.356 0.146 20.583 0.035 20.548 0.039 20.556 0.053 20.940 0.310 0.050 07.58 084.6 -69.4 sdA: 07836 00119 5.520 0.376 
000035.88-024622.11 4354-55810-0305 011 20.979 0.081 19.803 0.019 19.666 0.027 19.650 0.023 19.591 0.063 0.037 007.5 094.2 -62.8 sdA  07893 00065 6.143 0.189 07893 0065 5.940 0.190 0.153 0.0016
000043.52+351644.26 7145-56567-0818 006 21.132 0.112 20.585 0.030 20.531 0.040 20.534 0.051 20.477 0.175 0.071 015.3 111.4 -26.5 DA   09208 00104 8.095 0.149 09178 0104 7.820 0.150 0.521 0.064
000049.03-105805.58 7167-56604-0202 010 20.992 0.104 20.036 0.032 19.768 0.027 19.774 0.032 19.717 0.116 0.036 002.0 084.1 -69.9 sdA  07092 00083 6.180 0.259 07117 0083 6.003 0.260 0.149 0.0040
000052.60+265459.66 6511-56540-0042 017 19.480 0.044 18.577 0.022 18.258 0.015 18.174 0.017 18.129 0.028 0.046 005.7 109.2 -34.6 sdA  06862 00045 5.963 0.155 06889 0045 5.804 0.150 0.147 0.0009
000100.52-100222.12 7167-56604-0234 005 21.850 0.308 20.956 0.088 20.953 0.071 21.108 0.094 21.952 0.858 0.036 000.0 085.7 -69.2 DA   08977 00131 8.292 0.164 08969 0131 8.020 0.160 0.609 0.088
000110.91+285342.93 7134-56566-0368 004 21.433 0.116 21.032 0.041 20.916 0.045 20.958 0.069 20.535 0.172 0.068 00.00 109.9 -32.7 DA:  07722 00166 8.229 0.236 07735 0166 8.090 0.240 0.646 0.135
000133.32+170237.76 6173-56238-0428 005 21.173 0.121 20.752 0.038 20.974 0.059 21.087 0.078 22.163 0.567 0.028 000.0 106.2 -44.2 DA   11198 01000 7.690 0.500 11283 1000 7.540 0.500 0.409 0.176
000134.78+201514.44 6170-56240-0638 009 20.626 0.080 20.232 0.024 20.126 0.028 20.053 0.037 20.048 0.147 0.073 56.89 107.3 -41.1 DA   07512 00093 8.343 0.132 07520 0093 8.230 0.130 0.726 0.080
000134.86+321616.24 6498-56565-0910 009 20.541 0.059 20.204 0.025 20.184 0.035 20.428 0.062 20.858 0.314 0.051 000.0 110.8 -29.4 DA   10088 00089 8.189 0.094 10042 0089 7.920 0.090 0.558 0.041
\end{verbatim}
\end{minipage}
\end{table*}

\begin{table*}
\tiny
\noindent
\begin{minipage}{\linewidth}
\caption{New Classification of Known White Dwarf Stars.
\label{old}}
\begin{verbatim}
#SDSSJ              P-M-F           S/N u      su    g      sg    r      sr    i      si    z      sz    E(B-V) ppm  long  lat   sp   Teff  sT     logg slogg  Teff dTeff logg dlogg mass dmass
#                                       (mag) (mag) (mag)   (mag) (mag) (mag) (mag)   (mag) (mag) (mag) (mag)  0.001"  (o)   (o)     (K)   (K)  (cgs)  (cgs) (K)  (K) (cgs)  (cgs) (Msun)(Msun)
000054.40-090806.92 7167-56604-0806 022 19.317 0.042 18.997 0.033 18.952 0.019 19.030 0.029 19.094 0.061 0.050 53.63 087.0 -68.4 DQ  07957 01000 8.000 0.000 -log(C/He)=8.000 0.000
000307.06+241211.68 6879-56539-0704 064 16.159 0.022 16.190 0.022 16.556 0.028 16.806 0.017 17.053 0.026 0.147 04.41 109.0 -37.4 sdB 28566 00099 5.503 0.017
000309.26-060233.49 7147-56574-0956 007 20.571 0.058 20.065 0.021 19.970 0.020 19.966 0.027 19.926 0.074 0.041 054.6 092.2 -66.0 DA  08264 00097 7.157 0.218 08259 0097 6.940 0.220 0.234 0.039
000309.26-060233.49 7148-56591-0586 008 20.571 0.058 20.065 0.021 19.970 0.020 19.966 0.027 19.926 0.074 0.041 054.6 092.2 -66.0 DA  08489 00099 8.164 0.162 08500 0099 7.930 0.160 0.559 0.076
000321.60-015310.86 4365-55539-0502 015 19.232 0.028 19.216 0.027 19.294 0.020 19.498 0.025 19.728 0.072 0.041 053.8 096.4 -62.3 DA  06126 00103 9.320 0.214 06110 0103 9.370 0.210 1.285 0.072
\end{verbatim}
\end{minipage}
\end{table*}

\label{lastpage}
\end{document}